\documentclass[12pt, draftclsnofoot, onecolumn]{IEEEtran}
\IEEEoverridecommandlockouts
\usepackage{algpseudocode}
\usepackage{algorithm}
\usepackage[utf8]{inputenc}
\usepackage{xcolor}
\usepackage{color}
\usepackage{bm}
\usepackage{amsfonts}
\usepackage{amssymb}
\usepackage{amscd}
\usepackage{graphics}
\usepackage[dvips]{graphicx}
\usepackage{epsfig}
\usepackage{subfigure}
\usepackage{caption}
\usepackage[cmex10]{amsmath}
\usepackage{array}
\usepackage{eqparbox}
\usepackage{flushend}
\usepackage{hyperref}
\usepackage{fancyhdr}
\usepackage{titling}
\usepackage{enumerate}
\usepackage{stackrel}
\usepackage{pdflscape}
\usepackage{longtable}
\usepackage{lineno}
\usepackage{titling}

\newcommand{\abs}[1]{\left\vert#1\right\vert}
\newcommand{\norm}[1]{\Vert#1\Vert}

\makeatletter
\def\blfootnote{\xdef\@thefnmark{}\@footnotetext}
\makeatother

\newtheorem{theorem}{Theorem}[section]

\newtheorem{corollary}[theorem]{Corollary}

\newenvironment{definition}[1][Definition]{\begin{trivlist}
\item[\hskip \labelsep {\bfseries #1}]}{\end{trivlist}}

\newcommand{\qed}{\nobreak \ifvmode \relax \else
      \ifdim\lastskip<1.5em \hskip-\lastskip
      \hskip1.5em plus0em minus0.5em \fi \nobreak
      \vrule height0.75em width0.5em depth0.25em\fi}

\def\BibTeX{{\rm B\kern-.05em{\sc i\kern-.025em b}\kern-.08em
    t\kern-.1667em\lower.7ex\hbox{E}\kern-.125emX}}

\usepackage{newlfont}
\hypersetup{pageanchor=false}
\begin{document}
\title{Learning to Cache: Distributed Coded Caching in a Cellular Network With Correlated Demands}

\author{\IEEEauthorblockN{S. Krishnendu, B. N. Bharath, Navneet Garg, Vimal Bhatia  \\
		and Tharmalingam Ratnarajah}\thanks{S. Krishnendu and Vimal Bhatia are with Indian Institute of Technology Indore, India, e-mail: \texttt{\{phd1701102001,vbhatia\}@iiti.ac.in}. B. N. Bharath is with Indian Institute of Technology Dharwad, India, e-mail: \texttt{bharathbn@iitdh.ac.in}. Navneet Garg and Tharmalingam Ratnarajah are with Institute for Digital Communications, The University of Edinburgh, Edinburgh, U.K. e-mail: \texttt{navneet.garg4@gmail.com, T.Ratnarajah@ed.ac.uk.}}}

	 
\date{}
\maketitle

\begin{abstract}
Design of distributed caching mechanisms is considered as an active area of research due to its promising solution in reducing data load in the backhaul link of a cellular network. In this paper, the problem of distributed content caching in a small-cell Base Stations (sBSs) wireless network that maximizes the cache hit performance is considered. Most of the existing works focus on static demands, however, here, data at each sBS is considered to be correlated across time and sBSs. The caching strategy is assumed to be a weighted combination of past caching strategies. A high probability generalization guarantees on the performance of the proposed caching strategy is derived. The theoretical guarantee provides following insights on obtaining the caching strategy: (i) run regret minimization at each sBS to obtain a sequence of caching strategies across time, and (ii) maximize an estimate of the bound to obtain a set of weights for the caching strategy which depends on the discrepancy. Also, theoretical guarantee on the performance of the LRFU caching strategy is derived. Further, federated learning based heuristic caching algorithm is also proposed. Finally, it is shown through simulations using Movie Lens dataset that the proposed algorithm significantly outperforms LRFU algorithm.

\end{abstract}


\section{Introduction} \label{sec:intorduction}
In the recent past, there is a pressing need for revamping of the next generation wireless infrastructure network due to an unprecedented increase in the data demand \cite{Furuskar2015}. There has been several proposals for new wireless network designs towards alleviating the data demand problem. A few example designs include Fog network \cite{intharawijitr2016analysis} with edge computing, deployment of small cells to offload wireless data from a macro Base Station (BS), integrating existing WiFi access points to share the load, to name a few \cite{Bennis2013}, \cite{Chou2014}. It is well known that small-cell infrastructure with edge computing facility alone cannot support the data demand since the data clogging in the backhaul acts as a bottleneck. A new paradigm to handle this data clogging is through caching in the cellular networks. Caching can reduce the peak traffic by prefetching popular contents into memories at the small-cell Base Stations (sBSs) \cite{Niesen2012,Wu2016, 8755463}. Past works in caching include the classical work from the point-of-view of information theory by Neisen et al. \cite{info} (also, see \cite{borst}), combinatorial optimization approach \cite{Ji2016}, energy efficient caching of files in a Device-to-Device (D2D) network (see \cite{Ji2016} - \nocite{Zhang2016}\cite{Chen2018a}), and proactive caching strategy, as in \cite{Gregori2016}.

One of the key problems to be addressed in caching is that of estimating/predicting the popularity profile or demands of the files. Majority of the existing work assume static demands, and hence algorithms are designed to get a good estimate of the popularity profile (see \cite{Golrezaei2013}-\nocite{Tatar2014}\nocite{Bharath2016}\nocite{Song2017}\cite{Chen2016}). On the other hand, estimating the popularity profile based on the data assumes a naive estimate, i.e., a simple averaging, which may not perform well in highly non-stationary environments. However, the demands in reality are non-stationary, and perhaps correlated across time; this makes the algorithms designed for static demands/popularity profiles to under perform. A line of attack for solving this issue is to consider online learning algorithm to proactively cache the contents \cite{Gregori2016}. The authors in \cite{basu2018adaptive} showed that a good hit rate under non-stationary demands can be achieved through a Time to Live (TTL) based algorithm. Some past work assumed that there is a stationary caching policy such as Least Recently Used (LRU) \cite{gast}, CLIMB \cite{Starobinski, Coffman}, and k-LRU \cite{martina} and have characterized the learning errors as a function of time. The learning error depends on the stationary distribution, which in turn depends on the mixing time \cite{li2018accurate}. Many of these works result in a regret of $\Omega(T)$. A Multi-Arm-Bandit (MAB) approach to caching is considered in \cite{blasco2014learning}, and the authors show that a regret of $\mathcal{O}(\log T)$ can be achieved.

The approach taken so far is either online learning in the adversarial setting leading to regret minimization or by designing caching strategies by estimating the popularity profile (see \cite{ngarg}). The disadvantage in the adversarial setting is that the statistical pattern in the data is completely ignored. An improvement on this to account for statistical pattern is to combine the strategies in a systematic way, this is termed as online-to-batch conversion in the literature \cite{agarwal2012generalization}. There are several heuristics such as LRU, Least Frequently Used (LFU) and Least Recently Frequently Used (LRFU) (and its variants) which tend to work well in a non-stationary environment. These lack theoretical guarantees when the demand statistics are non-stationary. Therefore, in this paper, a systematic approach driven by theory to designing caching strategy when the demands/requests are highly \emph{non-stationary} will be addressed. Further, the mathematical tools developed are used to provide guarantees for LRFU, and its variants under non-stationary and correlated demands. 

In this paper, the problem of distributed caching across multiple sBSs with correlated demands across time as well as sBSs is considered. Since the demands can be correlated, a conditional average of the cache hit is considered as a metric to design caching strategies. Here, conditioning is with respect to the ``local" data available at the sBS. Following are the main contributions of this paper:
\begin{itemize}
\item  In this paper, coded caching is considered where a fraction of the file can be cached, and hence the caching strategy refers to a rule that assigns fraction of files to the sBS satisfying the cache size constraint. Designing a general optimal caching strategy without any structural assumptions is difficult. Therefore, a structure on the caching strategy is assumed, and a high probability guarantee/bound on the conditional average cache hit is derived using Martingale difference equation $\cite{zbMATH00044889}$. The structure imposed on the caching strategy is the following: the caching strategy employed by a sBS $b$ at time slot $t$ is a weighted combination of a sequence of caching strategies across time until time $t-1$, and the neighboring sBSs' caching strategies. The weights, termed here as \emph{caching-weights}, take into account the correlation across time as well as sBSs. The insights provided by the bound is used to design the caching algorithm, i.e., the caching-weights as well as the caching strategies across time.  
\item The derived bound in Theorem \ref{thm:mainresult_codedcaching} is shown to depend on several terms such as regret, discrepancy (a measure of correlation across time and sBSs), and weighted average of the past cache hits. Each of these in turn depends on the caching-weights. A part of the bound indicates that the caching strategies should be chosen in such a way that it minimizes the regret term. After solving regret minimization problem, the caching-weights are optimized to obtain the overall caching strategy. This objective function involves an estimate of the discrepancies, and a linear combination of the past cache hits. An iterative algorithm to solve this problem is proposed (see Section~\ref{sec:alg_iterative}). It is shown that this method results in better performance as compared to LRFU and equal weights. As a corollary of the main result, a guarantee on the performance of an algorithm using equal caching-weights is also obtained. It is shown that under independent and identically distributed (i.i.d.) demands, the performance of the proposed algorithm, LRFU, and equal caching-weights are similar. 
\item Using the mathematical tools developed in the paper, a theoretical guarantee on the performance of the LRFU caching strategy under non-stationary demands is derived. This guarantee is in terms of discrepancy terms that measures the statistical relationship between the demands at different time instants. Further, in the i.i.d. setting, it is shown that the LRFU performs close to the ``optimum" caching strategy, as expected. 
\item Motivated by the work in \cite{wang}, a federated learning based heuristic caching algorithm is proposed, where instead of optimizing local cache hit, a proximal term is added that takes care of the closeness between the local caching strategy and the average of the caching strategies of the neighboring sBSs. In doing so, the local solution, although tuned to its demands, will always be in close proximity to the average strategy. Numerical results show that the proposed algorithms (both federated caching and weighted averaging caching algorithm) significantly outperform (of the order of $10^4$ to $10^5$ cache hit improvement) LRFU as well as the equal weight algorithms. Further, the federated caching algorithm performs better than the weighted caching algorithm motivating our future work on proving guarantees on this heuristic algorithm. Finally, several useful insights and future directions are provided in Section~\ref{sec:conclusion}. 
\end{itemize}

\section{System Model} \label{sec:sys_model}
\begin{figure}[h!]
    \centering
    \includegraphics[width=20em,height=15em]{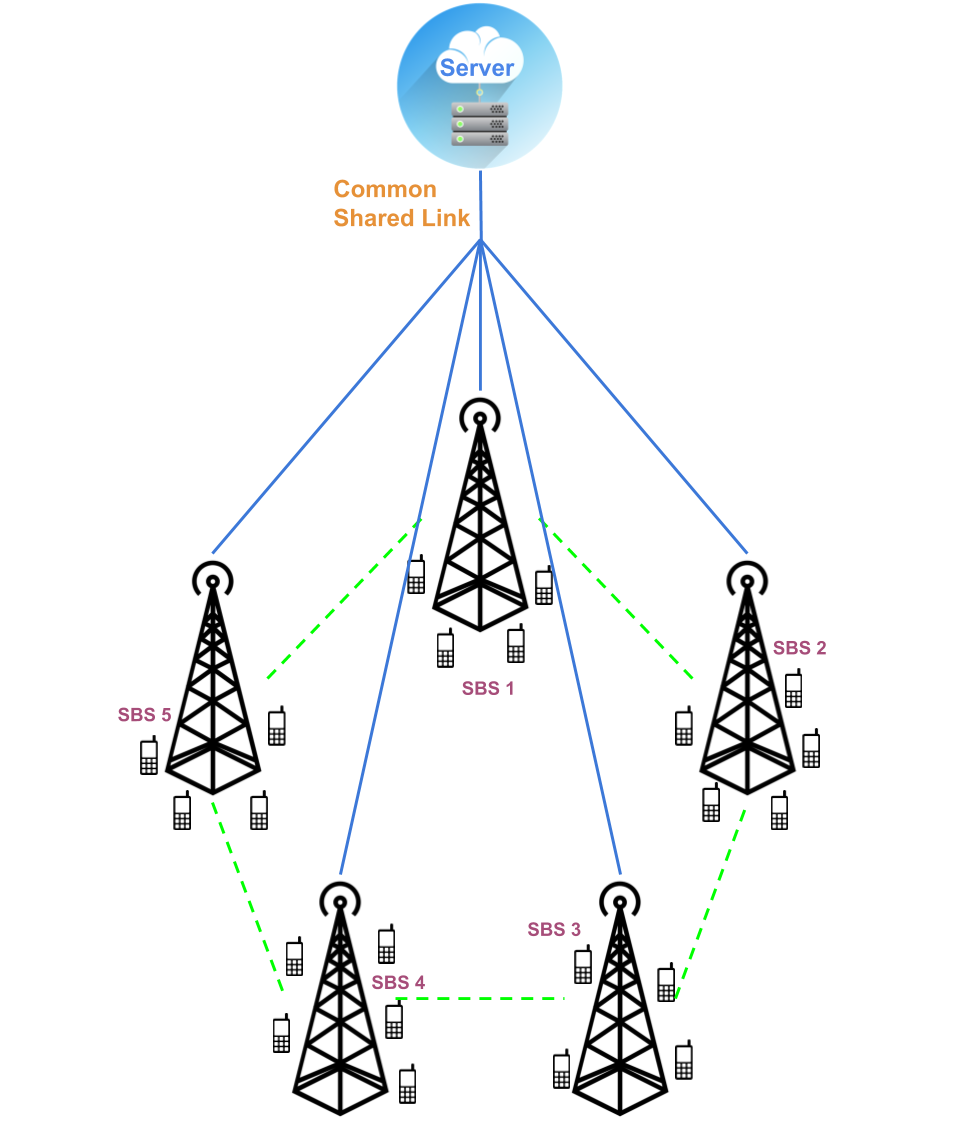}
    \caption{System model showing multiple sBSs connected to users with limited cache memory.}
    \label{fig:caching_EC}
\end{figure}

The system model consists of a cellular network with $M$ sBSs denoted by the set $\mathbb{B}$, and users denoted by the set $\mathbb{U}$, as shown in Fig.~\ref{fig:caching_EC}. Each sBS is assumed to have a limited computation facility and a cache memory of size $C$ bits to store popular contents. This computation capability facilitates distributed caching decisions to be taken at individual sBS without leveraging heavily on the central computing facility such as cloud service, thus saving tremendously on communication and computation costs. Further, it is assumed that the sBSs can communicate with each other through a limited capacity links. For example, the neighboring sBSs can share limited information such as caching decisions, popular demands and its trends amongst each other. Note that this edge computing paradigm with communication links between sBSs encompasses the proposed Fog network architecture \cite{intharawijitr2016analysis}. We assume a time slotted system, where in each slot a user requests contents from the content library $\mathcal{F}$ having $N$ contents, i.e., $\abs{\mathcal{F}} = N$. The demand for the content $f \in \mathcal{F}$ by the user $u$ in the slot $t$ is denoted by $d_{f,u}(t)$. The requests across time slots and sBSs can be correlated with an arbitrary distribution. Since in a practical content library, the files are of different sizes, hence the same is assumed in this work (see next subsection).

In the standard cellular network setting without caching, the requested file is served by the sBS to which the user is associated by fetching the content from the server through backhaul and front-haul links of the network. Note that in the current implementation, each user is associated with a single sBS based on the SINR criterion. Keeping minimal changes to the current design, it is assumed that the scheduler associates a user to a sBS based on the SINR criterion. Let the set of users associated to the sBS $b$ in the time slot $t$ be denoted by $\mathbb{U}_b(t)$. The total demand for the file $f$ at the sBS $b$ in the time slot $t$ is given by $D_{f,b}(t) = \sum_{u \in \mathbb{U}_b(t)} d_{f,u}(t)$. Let the data available at the sBS $b$ at time $T$ be denoted by $Z_{b,1}^T \subseteq \mathcal{Z}_{b,1}^T$, which includes demands of sBS $b$ until time slot $T$, and the data shared by the neighboring sBSs. Here, $\mathcal{Z}_{b,1}^{t}$ denotes the set of all possible demands and caching strategy of the neighboring sBSs at the end of time slot $t$. The exact data that the neighboring sBSs provide will be explained in the later part of this paper. Further, $Z_{G,1}^T := \bigcup_{b \in \mathbb{B}} Z_{b,1}^T$ denotes the global data till time $T$. The following subsections describe the caching strategy employed, and the corresponding metric used to find the optimal strategy. 
\subsection{Caching Policy} \label{subsec:sysmodel_caching_policy}
At each SBS $b$, the cache placement is assumed to happen at the end of every time slot. In this paper, a distributed caching policy is considered, i.e., at the end of time slot $t-1$ for each file $f$, the caching policy for the next time slot is given by $\pi_{b,f,t}: \mathcal{Z}_{b,1}^{t-1} \rightarrow \mathcal{C}_{b,f}$. Thus, the overall caching policy is defined as $\bm{\pi}_{b,t} := \times_{f=1}^N \pi_{b,f,t}: \mathcal{Z}_{b,1}^{t-1}\rightarrow \times_{f=1}^N \mathcal{C}_{b,f}$. The choice of $\mathcal{C}_{b,f}$ depends on the type of caching employed. Here, coded caching is employed, as explained below:
\begin{itemize}
\item \texttt{Coded caching:} In a typical coded caching scheme, an original file $f$ of size $K_f$ bits is mapped into $S_f$ sub-packets of size $l$ bits each in such a way that if a user recovers any $L_f$ out of $S_f$ sub-packets, it can recover the whole file. This gives the flexibility to store $L_f$ or less number of packets at each sBS, and the remaining packets can be fetched from the server. For the sake of simplicity in notation, $L_f$ is used to represent the number of packets instead of the size of the file in bits as in the case of uncoded caching scheme. Although storing any fraction is not possible, choosing $\mathcal{C}_{b,f} = [0,1]$ is a good approximation when the number of sub-packets, i.e, $L_f$ is large. Note that the caching strategy $\bm{\pi}_{b,t}$ is a vector of dimension $N$. Since the cache size is limited to $C$ bits, it imposes the constraint that $\sum_{f} \pi_{b,f,t} L_f l \leq C$. Here, $L_f l$ is the total number of bits that needs to be recovered under the coded caching scheme, and $\pi_{b,f,t}$ is the fraction of the packets stored. 
\end{itemize} 

The following subsection presents the problem of caching  addressed in this paper. 
\vspace{-0.7cm}
\subsection{Problem Statement} \label{subsec:sysmodel_problem_stat}
In either coded or uncoded caching scheme, the ``amount" of requests that are present in the caches of sBSs to which the users are connected is a good measure of performance; this is termed as \emph{hit rate}. In view of this, the hit rate at the sBS $b$ is given by
\begin{equation} \label{eq:hitrate_bsb}
\mathcal{R}_{b,t}(\bm{\pi_b}) := \sum_{f=1}^N \sum_{u \in \mathbb{U}_b(t)} d_{f,u}(t) \pi_{b,f} \mathcal{L}_f.
\end{equation} 
The above corresponds to the \emph{instantaneous} hit rate at the sBS $b$ in the time slot $t$ when coded caching strategy $\pi_{b,f} \in [0,1]$ is employed with $\mathcal{L}_f:=L_f l$. Note that the factor $l$ does not impact the structure of the solution, and hence omitted from the definition of the hit rate. Since the hit rate is random, a widely used measure of performance is the average cache hit, i.e., $\sum_{b}\mathbb{E}\{\mathcal{R}_{b,t}(\bm{\pi}_{b,t})\}$ \footnote{$\mathbb{E[\cdot ]}$ represents the statistical expectation operator.}, where the average is with respect to the global demands.\footnote{Note that the demands across sBSs as well as time slots are correlated. Hence, the expectation should be with respect to all the total randomness.}
 However, at time $t$, the sBS $b$ will have access to its ``local'' data $Z_{b,1}^{t-1}$, and hence, conditional mean is the appropriate metric, i.e., $\sum_{b}\mathbb{E}\{\mathcal{R}_{b,t}(\bm{\pi}_{b,t})\left \vert \right. Z_{b,1}^{t-1}\}$, where the expectation is conditioned on the local demands, i.e., $Z_{b,1}^T$. Thus, the following problem needs to be solved
\begin{eqnarray} \label{eq:the_problem}
&&\max_{\bm{\pi}_{b,t}} {\sum_{b}\mathbb{E}\{\mathcal{R}_{b,t}(\bm{\pi}_{b,t})\left \vert \right. Z_{b,1}^{t-1}\}} \nonumber \\
&&\text{subject to} \sum_{f} \pi_{b,f,t} \mathcal{L}_f  \leq C.
\end{eqnarray}
Let the set of all caching strategy be denoted by $\mathcal{C}:=\{\pi_{b,f}: \pi_{b,f} \geq 0, \sum_{f} \pi_{b,f} \mathcal{L}_f  \leq C\}$. The above is similar to the formulation considered in the prediction problems \cite{kuznetsov2016time}. Unfortunately, in the real world scenario, the conditional expectation is difficult to compute, and hence the above problem cannot be solved. One possible approach could be to estimate the conditional expectation, and use it as a proxy in the above problem. Since the user demands arrive in real-time, this estimate could be updated online. However, in this paper, instead of updating the estimates online, the solution for caching problem will be obtained online using the available ``local'' data. In the following sections, solution to the above problem for coded caching scenarios is presented. 

\section{Online Distributed Coded Caching} \label{sec:online_coded_caching}
Towards addressing the problem, a few structural assumptions are made on the caching strategy employed. In a typical online learning with adversarial framework, a natural metric to consider is the ``regret". In the present setting, the demands are random in nature and this corresponds to a stochastic setting rather than an adversarial setting, i.e., the nature reacts in a random fashion rather than an adversarial fashion. A well known strategy to handle this is through online-to-batch conversion \cite{mohri2018foundations}, which is as follows: (i) at time slot $t$, solve the regret minimization problem to get a sequence of caching strategies, and (ii) use the average of these caching strategies at time $t$. This has the advantage of providing $\mathcal{O}(\frac{1}{T})$ regret when the problem is stochastic. The model considered in this paper has added complexity that the demands of any sBS $b$ across time slots can be correlated. Further, it can also be correlated with the demands of other sBSs. In this scenario, a natural extension of online-to-batch conversion is to take the average of regret minimizing caching strategies across time as well as the sBSs \cite{agarwal2012generalization,kuznetsov2016time}. Towards this, consider the following weighted average of a sequence of caching strategies $\bm{\pi}_{b,t}$ from time slot $t= T-\tau + 1$ to $T$ given by 
\begin{equation}\label{eq:avg_caching}
 \bm{\bm{\bar{\pi}}}_{b,T+1} := \sum_{t=T-{\tau}+1}^T \alpha_{b,t}\bm{\pi}_{b,t},
\end{equation}
where $\alpha_{b,t}$'s are the non-negative weights that satisfy $\sum_{t = T-\tau + 1}^T \alpha_{b,t} = 1$. The symbol $\bm{\alpha}_{b,T} := (\alpha_{b,T-\tau + 1},\ldots,\alpha_{b,t})$ is used to denote vector of weights corresponding to the sBS $b$ from time slot $T-\tau + 1$ to $T$. Linear prediction model has been widely used in the literature due to its simplicity and effectiveness to solve problems and hence, the caching strategy has been taken as a weighted linear combination of all the neighboring SBSs caching strategies \cite{makhoul, singer, kozat}. 
It is important to note that the average of caching strategy across time is also a valid caching strategy, i.e., the set of all caching strategies $\mathcal{C}$ is a convex set. Since the demands are correlated across sBSs, a natural way to construct the caching strategy for the time slot $T+1$ is as follows
\begin{equation} \label{eq:pistart_coded_caching}
\bm{\pi}^{(av)}_{b,T+1} := w^{T+1}_{b} \bm{\bar{\pi}}_{b,T+1}  + \sum_{b^{'} \in \mathcal{N}_b} w^{T+1}_{j_b(b^{'})} \bm{\bar{\pi}}_{b^{'},T+1},
\end{equation}   
where the map $j_b: \mathcal{N}_b \rightarrow \{1,2,\ldots,\abs{\mathcal{N}_b}\}$, and the weights are chosen to be non-negative with the constraint given by $\sum_{b^{'} \in \mathcal{N}_b} w^{T+1}_{j_b(b^{'})} + w^{T+1}_{b} = 1$ $\forall$ sBS $b$. Now, the problem is to choose weights in such a way that the average cache hit is maximized. One can expect that in order to solve this problem, any sBS $b \in \mathbb{B}$ at the end of time slot $T$ should have access to neighboring sBSs' data. 
In this paper, a formal approach to answer the above is detailed. Obviously the choice of the weights $w^{T+1}_{j_b(b^{'})}$ as well as $\alpha_{b,t}$ depend on how relevant (i) is its past caching decisions to the current demands, and (ii) caching decisions of neighboring sBSs are to the sBS $b$. These are captured through the following notions of mismatch and regret.
\begin{definition}[Definition (Mismatch):]
The mismatch between a sBS $b$ and its neighbor with weights $w^{T+1}_{j_b(b^{'})}$, $b^{'} \in \mathcal{N}_b$ is given by
\begin{equation}
\texttt{M}_{b,T+1}({\bm{w}}_{\neq b, T}) := \sum_{b^{'}  \in \mathcal{N}_b} w^{T+1}_{j_b(b^{'} )} \Delta_{b,T+1}(\bm{\alpha}_{b,T+1}, \bm{\alpha}_{b^{'},T+1}),
\end{equation}
where the weight vector $\bm{w}_{\neq b, T}: = (w^{T+1}_{j_b(b^{'})}: b^{'} \in \mathcal{N}_b)$, and
\begin{eqnarray} \nonumber
\Delta_{b,T+1}(\bm{\alpha}_{b,T+1}, \bm{\alpha}_{b^{'},T+1}) &:=& \mathbb{E}\{\mathcal{R}_{b,T+1}({\bm{\bar{\pi}}_{b^{'},T+1}})\left \vert \right. Z_{b,1}^{T}\} - \mathbb{E}\{\mathcal{R}_{b,T+1}{(\bm{\bar{\pi}}_{b,T+1})\left \vert \right. Z_{b,1}^{T}}\}.
\end{eqnarray}
\end{definition}

The above captures mismatch or discrepancy across sBSs, which will help us in determining the relevance of the neighboring sBSs' decisions. If the mismatch is small for a sBS $b$ essentially means that the neighboring sBSs strategy performs well on the sBS $b$. Similarly, to determine the relevant caching strategies across time to the current time slot, and to measure the performance, the two key tools are discrepancy across time and the regret, which are defined as follows.
\begin{definition}[Definition (Discrepancy):]
Given local information at the sBS $b$ with caching strategies $\bm{\pi}_{b,t}$ for $b \in \mathbb{B}$, $t=T-\tau + 1,\ldots,T$, the discrepancy at the end of time slot $T$ is defined by 
\begin{equation} \label{eq:disclg}
\mathbb{D}_{b,T}(\bm{\alpha}_{b,T}) : = \sup_{\bm{\pi}_{b,t}: t=T-\tau+1,\ldots,T} \abs{\sum_{t=T-\tau + 1}^T \alpha_{b,t}\Delta \bar{\mathcal{R}}_{T,t}(\bm{\pi}_{b,t})}.
\end{equation}
where $\Delta \bar{\mathcal{R}}_{T,t}(\bm{\pi}_{b,t}):=\mathbb{E}\{\mathcal{R}_{b,T+1}(\bm{\pi}_{b,t})\left \vert \right. Z_{b,1}^{T}\} - \mathbb{E}\{\mathcal{R}_{b,t+1}(\bm{\pi}_{b,t})\left \vert \right. Z_{b,1}^{t}\}$.
\end{definition}

\begin{definition}[Definition (Regret):]
The regret at the sBS $b$ at time $T$ with respect to a sequence of strategy $\bm{\pi}_{b,t}$ is defined as 
\begin{equation} \label{eq:regret}
\texttt{Reg}_{b,T,\tau}(\bm{\pi}_{b,t}) : = \sup_{\bm{\pi}_{b,t}^{*}}\sum_{t=T-\tau + 1}^T \mathcal{R}_{b,T+1}(\bm{\pi}_{b,t}^{*}) -\sum_{t=T-\tau + 1}^T \mathcal{R}_{b,T+1}(\bm{\pi}_{b,t}).
\end{equation}
\end{definition}
The following theorem gives guarantees for the proposed caching strategy, and also provides insights on how to choose the weights, and the sequence of caching policies across time.  The main result of the paper is stated below, and the corresponding proof is presented in Sec.~\ref{sec:proof_main_res_codedcaching}.
\begin{theorem} \label{thm:mainresult_codedcaching}
Given weights and a sequence of caching strategies as in \eqref{eq:pistart_coded_caching} that is adapted to $Z_{b,1}^T$, with a probability of at least $1-\delta$, $\delta > 0$, the following two bounds hold:
\begin{eqnarray} 
\mathbb{E}\left[\mathcal{R}_{b,T+1}(\bm{\bm{\pi_{b,T+1}}}^{(av)}) \left | \right. Z_{b,1}^T\right]  &\geq& \sum_{t=T-\tau + 1}^T\alpha_{b,t} \mathcal{R}_{b,t}(\bm{\pi}_{b,t}) - \mathcal{E}^{(1)}_{b,T}, \label{eq:azuma_At_thirdequation_thm}
\end{eqnarray}
where $$\mathcal{E}^{(1)}_{b,T}:=H_{\texttt{max}} \norm{\bm{\alpha}_{b,T}}_2 \sqrt{\frac{2}{\tau} \log\frac{1}{\delta}} + \texttt{M}_{b,T+1}(\bm{w}_{\neq b, T})  + \mathbb{D}_{b,T}(\bm{\alpha}_{b,T}),$$ $H_{\texttt{max}}$ is the maximum cache hit, and 
\begin{eqnarray}
 \mathbb{E}\left[\mathcal{R}_{b,T+1}(\bm{\pi_{b,T+1}}^{(av)}) \left | \right. Z_{b,1}^T\right]  &\hspace{-0.2cm}\geq\hspace{-0.2cm}& \sup_{\bm{\pi}_{b,t}} \sum_{t=T-\tau}^T \alpha_{b,t} \bar{\mathcal{R}}_{t,T}(\bm{\pi}_{b,t})  -  \mathcal{E}^{(2)}_{r,T}. \label{eq:azuma_At_sixthequation_thm}
\end{eqnarray}
for any $\gamma > 0$. In the above, $\bar{\mathcal{R}}_{t,T}(\bm{\pi}_{b,t}):=\mathbb{E}\left[\mathcal{R}_{b,T+1}(\bm{\pi}_{b,t})\left | \right. Z_{b,1}^T\right]$, and 
\begin{eqnarray}
\mathcal{E}_{b,T}^{(2)} &:=& 2H_{\texttt{max}} \norm{\bm{\alpha}_{b,T}}_2 \sqrt{\frac{2}{\tau} \log\frac{1}{\delta}} + \texttt{M}_{b,T+1}(\bm{w}_{\neq b, T}) +  \frac{2 \texttt{Reg}_{b,T,\tau}(\bm{\pi}_{b,t}) }{\tau}  \nonumber \\ &+& H_{\texttt{max}} \sum_{t=T-\tau + 1}^T \abs{{\alpha}_{b,t} - \frac{1}{\tau}} + 2 \mathbb{D}_{b,T}(\bm{\alpha}_{b.T}) + \gamma.
\end{eqnarray}
\end{theorem}

An important special case of the above result is when uniform caching strategy is used, i.e., $\alpha_{b,t} = 1/\tau$ $\forall$ $t$, which is presented as a corollary.
\begin{corollary} \label{corr:main_result}
Given equal weights, i.e., $\alpha_{b,t} = 1/\tau$ $\forall$ $t$, and a sequence of caching strategies as in \eqref{eq:pistart_coded_caching} that is adapted to $Z_{b,1}^T$, with a probability of at least $1-\delta$, $\delta > 0$, the following two bounds hold:
\begin{eqnarray} 
\mathbb{E}\left[\mathcal{R}_{b,T+1}(\bm{\bm{\pi_{b,T+1}}}^{(av)}) \left | \right. Z_{b,1}^T\right]  &\geq& \frac{1}{\tau}\sum_{t=T-\tau + 1}^T\mathcal{R}_{b,t}(\bm{\pi}_{b,t}) - \mathcal{E}^{(1)}_{b,T}, \label{eq:azuma_At_thirdequation_corr}
\end{eqnarray}
where $$\mathcal{E}^{(1)}_{b,T}:=\frac{H_{\texttt{max}}}{\tau}  \sqrt{2 \log\frac{1}{\delta}} + \texttt{M}_{b,T+1}(\bm{w}_{\neq b, T})  + \mathbb{D}_{b,T}(\bm{u}_{\tau}),$$ $H_{\texttt{max}}$ is the maximum cache hit, and 
\begin{eqnarray}
 \mathbb{E}\left[\mathcal{R}_{b,T+1}(\bm{\pi_{b,T+1}}^{(av)}) \left | \right. Z_{b,1}^T\right]  & \geq& \sup_{\bm{\pi}_{b,t}} \frac{1}{\tau} \sum_{t=T-\tau}^T \bar{\mathcal{R}}_{t,T}(\bm{\pi}_{b,t})  -  \mathcal{E}^{(2)}_{r,T}. \label{eq:azuma_At_sixthequation_corr}
\end{eqnarray}
for any $\gamma > 0$. In the above, $\bm{u}_{\tau}: = (\frac{1}{\tau},\frac{1}{\tau},\ldots,\frac{1}{\tau}) \in \mathbb{R}^{1 \times \tau}$, $\bar{\mathcal{R}}_{t,T}:=\mathbb{E}\left[\mathcal{R}_{b,T+1}(\bm{\pi}_{b,t})\left | \right. Z_{b,1}^T\right]$, and $\mathcal{E}_{r,T}^{(2)}:=\frac{2H_{\texttt{max}}}{\tau} \sqrt{2 \log\frac{1}{\delta}} + \texttt{M}_{b,T+1}(\bm{w}_{\neq b, T}) +  \frac{2 \texttt{Reg}_{b,T,\tau}(\bm{\pi}_{b,t}) }{\tau}  + 2 \mathbb{D}_{b,T}(\bm{u}_{\tau}) + \gamma$.
\end{corollary}

A few observations are in order with reference to Theorem \ref{thm:mainresult_codedcaching}. The term $H_{\texttt{max}} \sum_{t=T-\tau}^T \abs{\alpha_{b,t} - \frac{1}{\tau}}$ in the second bound suggests that all the weights should be close to $1/\tau$, i.e., uniform weights. On the other hand, both the bounds also suggest that the discrepancies should be made low by choosing the weights appropriately. This requires non-uniform weights in general. Since the two tasks are conflicting, a nice balance needs to be maintained by properly choosing the weights. Further, it is clear from the second bound that the caching policy should be chosen in such a way that the regret is minimized. The following subsection presents a systematic approach to find an online distributed caching algorithm.

\subsection{Algorithm for Online Distributed Coded Caching} \label{sec:alg_iterative}
In this subsection, the insights provided by the theory is used to propose an algorithm for distributed online caching. The main result states that upon using the caching strategy given in \eqref{eq:pistart_coded_caching}, the resulting cache hit is lower bounded by the expression in \eqref{eq:azuma_At_sixthequation_thm} with high probability. Now, at time slot $T+1$, the goal is to choose the individual strategy $\bm{\pi}_{b,t}$ to construct $\bm{\pi}^{(av)}_{b,T+1}$ as in \eqref{eq:pistart_coded_caching} such that the right hand side of  \eqref{eq:azuma_At_sixthequation_thm} consisting of regret and discrepancy terms to be maximized.\footnote{The regret and discrepancy have negative signs on the right hand side.} In particular, this can be done by using the following two steps: (i) choose the sequence $\bm{\pi}_{b,t}$ in such a way that the regret term is minimized, and (ii) minimize the mismatch terms $\texttt{M}_{b,T+1}(\bm{w}_{\neq b, T})$ and  $\mathbb{D}_{b,T}(\bm{\alpha}_{b,T})$ to get the optimal weights, which can be used to combine the caching sequence as in \eqref{eq:pistart_coded_caching}. The first step would be to find the regret minimizing caching strategy by solving the following optimization problem
\begin{eqnarray} \label{eq:regret_min_problem}
\min_{\bm{\pi}_{b,t}: \bm{1}^T \bm{\pi}_{b,t} \leq C} \left[\sup_{\bm{\pi}_{b,t}^{*}}\sum_{t=T-\tau + 1}^T \mathcal{R}_{b,T+1}(\bm{\pi}_{b,t}^{*}) -\sum_{t=T-\tau + 1}^T \mathcal{R}_{b,T+1}(\bm{\pi}_{b,t})\right]
\end{eqnarray}
to get a sequence of caching policies denoted by $\bm{\pi}^R_{b,t}$ $\forall$ $t$. Note that the above problem can be solved optimally at the end of time slot $T$ as each sBS has access to the demands until time slot $T$. Next step would be to maximize the right hand side of \eqref{eq:azuma_At_sixthequation_thm} excluding the regret term. Unfortunately, the discrepancy term is unknown, and hence is estimated using the demands. Moreover, the discrepancy term involves an optimization. One way to deal with this is to use the regret minimizing caching strategy, and solve the following optimization problem to obtain the weights
\begin{eqnarray} \label{eq:disc_opt}
&\stackrel{\sup}{\alpha_{b,t}, \bm{w}_{\neq b, T}}& \sum_{t=T-\tau + 1}^T \alpha_{b,t} \mathcal{R}_{b,t}(\bm{\pi}^R_{b,t}) - a \widehat{\mathbb{D}_{b,T}(\bm{\alpha}_b)} - \nonumber\\ 
&& b \widehat{\texttt{M}}_{b,T+1}(\bm{w}_{\neq b, T}) + \lambda \sum_{t=T-\tau + 1}^T \abs{\alpha_{b,t} - \frac{1}{\tau}}
\end{eqnarray}
for some $\lambda > 0$, and $\widehat{\mathbb{D}_{b,T}(\bm{\alpha}_b)}$ is an estimate of the discrepancy given by 
\begin{equation} \label{eq:disc_estimate}
\widehat{\mathbb{D}_{b,T}(\bm{\alpha}_b)}: = \sup_{\bm{\pi}_{b,t}: \bm{\pi}_{b,t}\bm{1}^T x \leq C} \abs{\sum_{t=T-\tau + 1}^T \alpha_{b,t} \sum_{f} \psi_{\tau_1,\tau_2}^{(f)}(t,T) \pi_{b,f}(t)},
\end{equation}
where $\psi_{\tau_1,\tau_2}^{(f)}(t,T):= \mathcal{L}_f \left(\frac{1}{\tau_1} \sum_{l=T-\tau_1 + 1}^T \phi_{b,f}(l) - \right.$ $\left. \frac{1}{\tau_2} \sum_{l=t-\tau_2 + 1}^{t-1} \phi_{b,f}(l)\right)$, and the sum demand $\Phi_{b,f}(t) :=  \sum_{u \in \mathbb{U}_b(t)} d_{f,u}(t)$. The constants $a$ and $b$ are fine tuned to get better results. An estimate of the discrepancy across sBSs is given by 
\begin{eqnarray}
   \widehat{\texttt{M}}_{b,T+1}(\bm{w}_{\neq b, T})&:=& \frac{1}{\tau} \sum_{b^{'} \in \mathcal{N}_b} w_{j_b(b^{'})}^{T+1} \left[\sum_{s,l=T-\tau + 1}^T \alpha_{b,s}\mathcal{R}_{b,l}(\pi^R_{b,s})
    - \right. \nonumber \\
    &\hspace{-0.6cm}&\left. \sum_{s,l=T-\tau + 1}^T \alpha_{b^{'},s}\mathcal{R}_{b^{'},l}(\pi^R_{b^{'},s})\right]. 
\end{eqnarray}
Note that the conditional expectations are replaced by the time average of the cache hit as a proxy to get the above estimate of the discrepancy. In the time slot $T$, the average cache hit from the time slot $T-\tau + 1$ to $T$ is used as a proxy for the conditional mean in the expression for $\texttt{M}_{b,T+1}(\bm{w}_{\neq b, T})$. Although the objective in \eqref{eq:disc_opt} seems to be simple, it is a non-convex function of $\alpha_{b,t}$ and $\bm{w}_{\neq b, T}$, making the problem difficult to solve for global optima. However, a simple gradient descent algorithm can be used to achieve a local optima. Using the gradient descent approach leads to \textbf{Algorithm $1$}, which is explained next. Note that the estimate of discrepancy above involves solving an optimization problem with respect to the caching strategy $\bm{\pi}_{b,t}$. However, this optimization problem depends on  ${\alpha}_{b,t}$, which is unknown. A natural approach to this is to assume some initial ${\alpha}_{b,t}$, and solving the above optimization problem using gradient descent step, and project to satisfy the cache constraint. This is done in steps $1$ and $2$ of the \textbf{Subroutine}. Using this, in the step $k+1$, an update $\pi^{t}_{b,f}(k+1) $ is obtained. This is used in the expression for an estimate of the discrepancy in \eqref{eq:disc_estimate}, and used in \eqref{eq:disc_opt} to subsequently solve for weights $\alpha_{b,t}$ and $w_{j_b(b^{'})}^{T+1}$. This is done by taking a gradient descent step with respect to $\alpha_{b,t}$ in the problem in \eqref{eq:disc_opt} followed by projection to satisfy the constraint $\sum_{t = T-\tau+1}^T \alpha_{b,t} = 1$. These two steps correspond to steps $3$ and $4$ of the \textbf{Subroutine}. Similar gradient steps are taken for the weights $w_{j_b(b^{'})}^{T+1}$. These steps correspond to steps $5$ and $6$ of the \textbf{Subroutine}. The details are provided in the algorithm below, and explained later in this section. 

\begin{algorithm}[h]
	
\caption{Cache Placement Algorithm}
\begin{algorithmic}[1]

\For{\text{$T=1,2,\ldots$, and sBS \; $b \in \mathbb{B}$}}
        \State \small{\text{Run regret minimization as in \eqref{eq:regret_min_problem} to get a sequence $\bm{\pi}^R_{b,t}$, $t = 1,\ldots,T$}}
 \State \text{Call \textbf{Subroutine ($T$, $\tau$, $\bm{\pi}^R_{b,t}$, $\bm{\pi}^R_{b^{'},t}$ for all $b^{'} \in \mathcal{N}_b$) to get $\bm{\pi}_{b,T+1}$}}. 
            
\EndFor

\end{algorithmic}
\end{algorithm}

The stopping criterion of the algorithm in the \textbf{Subroutine} is determined by checking if the difference in weights is smaller than a threshold. The threshold is chosen based on extensive simulations. The learning rate $\eta_k$, $\beta_k$, and $\gamma_k$ are chosen such that it decays as $1/\sqrt{k}$ with the iteration $k$. 

\vspace{0.4cm}	
\hrule
\vspace{0.1cm}	
\hrule
\textbf{Subroutine ($T$, $\tau$, $\bm{\pi}^R_{b,t}$, $\bm{\pi}^R_{b^{'},t}$ for all $b^{'} \in \mathcal{N}_b$):}
\vspace{0.1cm}
\hrule
\vspace{0.1cm}
\hrule

\begin{itemize}
\item \textbf{for} each sBS $b$, \textbf{for} $k=0,1,2,\ldots$ \textbf{do}
\begin{enumerate}
\item \textbf{If ($k=0$)}, then \texttt{initialize} $\pi_{b,f}^{(0)} = \frac{C}{\sum_{f} \mathcal{L}_f}$, $\forall$ $f$ and $\alpha_{b,t}^{(0)} = 1/\tau, \forall t \geq T- \tau + 1$, and zero otherwise. Let $\Gamma_{t,T} := \sum_{f}\pi_{b,f}^{t}(k)\Psi^{f}_{\tau_1,\tau_2}(t,T)$. For $k \neq 0$, \texttt{update}
\begin{equation}
     \pi^{t}_{b,f}(k+1) = \pi^{t}_{b,f}(k) + 2 \eta_{k} g, 
\end{equation}
where $g:=\alpha_{b,t}(k) \Psi^{f}_{\tau_1,\tau_2}(t,T)$ if $\sum_{t=T-\tau + 1}^T \Gamma_{t,T} > 0$, else choose $g:= -\alpha_{b,t}(k) \Psi^{f}_{\tau_1,\tau_2}(t,T)$.
\item \texttt{Project:} $\pi_{b,f}{(k+1)} \leftarrow \max\{\pi_{b,f}{(k+1)},0\}$ and $\pi_{b,f}^{(k+1)} \leftarrow \frac{C \pi_{b,f}{(k+1)}}{\sum_{f}{\pi_{b,f}{(k+1)} \mathcal{L}_f}}$.
\item \texttt{Update the $\alpha$-weights:}
\begin{eqnarray}
  \alpha_{b,t}(k+1) &=& \alpha_{b,t}(k) + \beta_k\left[\mathcal{R}_{b,t}(\pi^R_{b,t}) -  \Theta  \right. - \left. \nabla_\alpha \widehat{\texttt{M}}_{b,T+1}(\bm{w}_{\neq b, T})\right],
\end{eqnarray}
where $\beta_k$ is the step size, $\Gamma_{t,T}$ is as defined in step $1$ above, $\Theta:= 2\max\{\Gamma_{t,T}, -\Gamma_{t,T}\}
    -\lambda \nabla_\alpha \norm{{\alpha}_{b,t}(k) - u}_1$, $$\nabla_\alpha \widehat{\texttt{M}}_{b,T+1}(\bm{w}_{\neq b, T}) := \frac{2}{\tau} \sum_{b^{'}\in \mathcal{N}_b}w^{T+1}_{j_b(b^{'})}(k) \hspace{-0.2cm} \sum_{l=T-\tau + 1}^T \hspace{-0.2cm}  \mathcal{R}_{b,l}(\pi^R_{b,t}),$$ and $\nabla_\alpha \norm{{\alpha}_{b,t}(k) - u}_1: = \bm{1}\{\alpha_{b,t} < \frac{1}{\tau}\} - \bm{1}\{\alpha_{b,t} \geq \frac{1}{\tau}\}$. 
\item \texttt{Project:} $\alpha_{b,t}{(k+1)} \leftarrow \max\{\alpha_{b,t}{(k+1)},0\}$, and $\alpha_{b,t}{(k+1)} \leftarrow \frac{\alpha_{b,t}{(k+1)}}{\sum_{t= T-\tau +1}^T \alpha_{b,t}{(k+1)}}$.
\item \texttt{Update the $\bm{w}$-weights} for sBS $b$ using data from neighboring sBSs as follows:
\begin{eqnarray} \nonumber
     w^{T+1}_{j_b(b^{'})}(k+1) &=& w^{T+1}_{j_b(b^{'})}(k) - \frac{2\gamma_k}{\tau} \left[\sum_{s,l=T-\tau + 1}^T \alpha_{b,s}(k) \right. \nonumber \\ 
     &&\hspace{-3cm} \left. \times \mathcal{R}_{b,l}(\pi^R_{b,s})
    - \sum_{s,l=T-\tau + 1}^T \alpha_{b^{'},s}(k)\mathcal{R}_{b,l}(\pi^R_{b^{'},s})\right] 
\end{eqnarray}
\item \texttt{Project:} $w^{T+1}_{j_b(b^{'})}(k+1) \leftarrow \max\{w^{T+1}_{j_b(b^{'})}(k+1),0\}$, and 
\begin{itemize}
    \item \texttt{Normalize:} If $\sum_{b^{'} \in \mathcal{N}_b}w^{T+1}_{j_b(b^{'})}(k+1) < 1$, then
   $ w^{T+1}_{b} = 1-\sum_{b^{'} \in \mathcal{N}_b}w^{T+1}_{j_b(b^{'})}(k+1)$,
else $w^{T+1}_{b} = 0$ and for all $b^{'} \in \mathcal{N}_b$,
 $   w^{T+1}_{j(b^{'})}(k+1) = \frac{w^{T+1}_{j_b(b^{'})}(k+1)}{\sum_{b^{'} \in \mathcal{N}_b}w^{T+1}_{j_b(b^{'})}(k+1)}$.
\end{itemize}
\item \textbf{if (not converged):} Broadcast the weights obtained in the current iteration to all neighboring sBSs, and go back to step $1$ \textbf{else;} return
\begin{eqnarray}
      \bar{\bm{\pi}}_{b,T+1} &=& w_{b}^{T+1}(k+1)\sum_{t=T-\tau + 1}^T \alpha_{b,t}(k+1) \bm{\pi}^{R}_{b,t} + \nonumber \\
      && \hspace{-1.2cm} \sum_{b^{'} \in \mathcal{N}_b} w^{T+1}_{j_b(b^{'})}(k+1)\sum_{t=T-\tau + 1}^T \alpha_{b^{'},t}(k+1) \bm{\pi}^{R}_{b^{'},t}
\end{eqnarray}
\end{enumerate}
\item \textbf{end for}
\end{itemize}
\vspace{0.2cm}	
\hrule
\hrule
\vspace{0.4cm}	
Since the above algorithm is a modification of gradient descent algorithm,\footnote{The algorithm deviates from the classical gradient descent in the step $2$ of the subroutine as the problem involves two optimization problems.} the convergence can be proved in a similar manner to that of classical gradient descent. The proof is omitted due to lack of space. In the following subsection, a simple \\
 for caching mechanism design that takes into account neighboring SBSs requests is proposed.

\subsection{Federated Learning Based Heuristics Caching Mechanism} \label{sec:heuristic}
In the single SBS scenario, a natural approach to find a caching strategy is to solve the following optimization problem:
\begin{equation}
\min_{\pi: \sum_f \pi_f l_f \leq L} \hat{F}_k(\pi),
\end{equation}
where $\hat{F}_k(\pi) :=\sum_{f \in \mathcal{F}}(1- \pi_{f})l_f\hat{d}_{f,k}^{(t)}$ is an estimate of the average cache miss, and $\hat{d}_{f,k}^{(t)}:= \frac{1}{\tau} \sum_{s= t-\tau}^{t-1} d_{f,k}^{(s)}$. However, if the amount of data available is less, the estimate will be poor, and hence results in a poor caching strategy. One way to overcome this is to use the information available from the neighboring sBSs. This can be done by penalizing the caching strategies that are far from some average of the caching strategies of the neighboring sBSs, i.e., $\lambda \norm{\bm{\pi} - \bm{\bm{\bar{\pi}}}_{\mathcal{N}_k}^{(t)}}^2$, where $\bm{\bm{\bar{\pi}}}_{\mathcal{N}_k}^{(t)}$ is the average of neighboring SBSs caching strategies. This requires information requires past caching strategies from the neighboring sBSs, which is assumed to be available. The parameter $\lambda > 0$ controls the amount of deviation that can be tolerated. More details of the heuristic algorithm are provided in \textbf{Algorithm $2$}. The parameters in the Federated caching algorithm are fine tuned to get better performance. The following subsection presents an analysis of the LRFU scheme. To the best of authors knowledge, this analysis is the first of its kind in the literature. 

\begin{algorithm}
		\caption{Algorithm for Distributed Caching (Heuristics)}
		\begin{algorithmic}[1]
			\Procedure{Federated Learning for caching}{}
			\For{ $\forall$ sBS $k = 1,\ldots, N$ and $\forall f= 1, \ldots,F$}
			\State \text{$ \hat{d}_{f,k}^{(0)}$ $\gets$ initial demand }
			\State \text{ $\pi_{\mathcal{N}_k}^{(0)}  \gets$ initial caching vector	s.t. $\sum_{f} {\pi}_{f} \mathcal{L}_f \leq C$}
			\EndFor
			\For{$t=1,2\ldots,$}
			\State \text{sBS $k$ sents $\hat{\pi}_{k, t-1}^*$ to its neighboring sBSs.} 
			\State \text{At each sBS $k$, estimate demand vectors and average caching vectors as follows:}
		     \begin{equation}
		     \hat{d}_{f,k}^{(t)}:= \frac{1}{\tau} \sum_{s= t-\tau}^{t-1} d_{f,k}^{(s)}, \text{ and } \bm{\bar{\pi}}_{\mathcal{N}_k}^{(t)} := \frac{1}{|\mathcal{N}_k|} \sum_{ j \in \mathcal{N}_k} \hat{\pi}_{k, t-1}^*. 
		     \end{equation}
			\State \text{Solve the following optimization problem to get $\hat{\pi}_{k,t}^{*}$:}
			\begin{equation}
			\hat{\pi}_{k,t}^{*} := \arg \min_{\bm{\pi}} \hat{F}_{k,t}(\pi) + \lambda\norm{\bm{\pi} - \bm{\bm{\bar{\pi}}}_{\mathcal{N}_k}^{(t)}}^2,
			\end{equation}
			where $\hat{F}_{k,t}(\pi) := \sum_{f \in \mathcal{F}}(1- \pi_{f,k})\mathcal{L}_f\hat{d}_{f,k}^{(t)}$, and $\lambda > 0$. 
			\State \text{Cache files at sBS $k$ according to $\hat{\bm{\pi}}_{k,t}^*$, and distribute across its neighboring sBSs.}				
			\EndFor
			\EndProcedure
		\end{algorithmic}
		
	\end{algorithm}

\subsection{LRFU Caching Policy: Analysis and Guarantees} \label{sec:lrfu}
In this scheme, an average of the past demands of each file is listed in the decreasing order, and the first $k$ files are stored, where $k$ is chosen in such a way that the cache size constraint is satisfied. In particular, in time slot $t$, at sBS $b$, the following optimization problem is solved:
\begin{equation}
\max_{\bm{\pi}: \sum_{f} \pi_{f} \mathcal{L}_f \leq C} \sum_{f} {\pi}_{f} \hat{d}_{b,f,t} \mathcal{L}_f,
\end{equation}
where $\hat{d}_{b,f,t} := \frac{1}{\tau} \sum_{s=t-\tau - 1}^{t-1} d_{b,f,s}$ $\forall$ $f$. In the case of constant file sizes, i.e., $\mathcal{L}_f := L$ $\forall$ $f$, the solution to the above amounts to listing the files in the decreasing order of $\hat{d}_{b,f,t}$, and storing the top $k$ files, where $k$ is chosen to satisfy the cache constraint. However, when files sizes are different, instead of the ``average" demands $\hat{d}_{b,f,t}$, one should consider $\mathcal{L}_f \hat{d}_{b,f,t}$ in the above argument. By imposing the constraint that $\pi_{f} \in \{0,1\}$ $\forall$ $f$ leads to the classical LRFU solution. Let the corresponding caching strategy be denoted by $\bm{\pi}_{b,t}^{\texttt{LRFU}}$. Before stating the main theorem, the following notions of discrepancy (similar to discrepancy described earlier) will be used to state the main result.
\begin{definition}[Definition (Discrepancy across time and information):]
Given local and global information at the sBS $b$ with caching strategies $\bm{\pi}_{b,t}$ for $b \in \mathbb{B}$, $t=T-\tau + 1,\ldots,T$, the corresponding discrepancy between local and global information at the end of time slot $T$ is defined by 
\begin{equation} \label{eq:disclg_globallocal}
\mathbb{D}_{GL,T}(\tau) : = \sup_{\bm{\pi}_{b,t}: t=T-\tau+1,\ldots,T} \abs{\frac{1}{\tau}\sum_{t=T-\tau + 1}^T \left(\Delta \bar{\mathcal{R}}_{T,t}\right)},
\end{equation}
where $\Delta \bar{\mathcal{R}}_{T,t}:=\mathbb{E}\{\mathcal{R}_{b,T+1}(\bm{\pi}_{b,t})\left \vert \right. Z_{b,1}^{T}\} - \mathbb{E}\{\mathcal{R}_{b,t}(\bm{\pi}_{b,t})\left \vert \right. Z_{G,1}^{T}\}$.
\end{definition}

The above measures the discrepancy between the local and the global data, i.e., the demands at sBS $b$ and all other sBSs. In the i.i.d. demands scenario, it is clear that the discrepancy is zero, as expected. In other words, having access to global information is useful to improve the accuracy of the future demand estimate through averaging, and hence the average cache hit as well. The following theorem provides guarantees on the performance of the LRFU scheme in comparison with \eqref{eq:the_problem}, which assumes perfect knowledge of statistics of the demands. Note that the analysis used in the proof of the following result does not depend on whether $\pi_{f} \in \{0,1\}$ or $\pi_{f} \in [0,1]$. Therefore, this constraint is not explicitly stated. 
\begin{theorem} \label{thm:lrfu_guarantees}
For the LRFU caching strategy $\bm{\pi}_{b,t}^{\texttt{LRFU}}$, with a probability of at least $1-\delta$, $\delta > 0$, the following bound hold:
\begin{equation} 
{\displaystyle
\sum_{f} \bm{\pi}_{b,t}^{\texttt{LRFU}} \hat{d}_{b,f,t} \mathcal{L}_f \leq \sup_{\bm{\pi_{b}}} \mathbb{E}\left[\mathcal{R}_{b,t}(\bm{\pi_{b}}) \left | \right. Z_{b,1}^{t-1}\right] + \mathbb{D}_{GL,t}(\bm{\alpha}_b) + \mathbb{D}_{b,t}(\bm{u}_{\tau}) + H_{\texttt{max}} \norm{\bm{\alpha_{b,T}}}_2 \sqrt{\frac{2 \log{\frac{1}{\delta}}}{\tau}} } \label{eq:lrfu_guarantee} 
\end{equation}
where $\mathbb{D}_{b,t}(\bm{u}_{\tau})$ is as defined in \eqref{eq:disclg} with $\bm{u}_\tau := (\frac{1}{\tau},\frac{1}{\tau},\ldots,\frac{1}{\tau})$ is a $1 \times \tau$ vector, and $\mathbb{D}_{GL,t}(\bm{\alpha}_b) $ is as defined in \eqref{eq:disclg_globallocal}. 
\end{theorem}
\emph{Proof:} See Appendix \ref{sec:proof_main_res_LRFU}. 

It is clear from the above thoerem that in the i.i.d. demands scenario, the right hand side will be $\sup_{\bm{\pi_{b}}} \mathbb{E}\left[\mathcal{R}_{b,t}(\bm{\pi_{b}}) \left | \right. Z_{b,1}^{t-1}\right]  + H_{\texttt{max}} \norm{\bm{\alpha_{b,T}}}_2 \sqrt{\frac{2 \log{\frac{1}{\delta}}}{\tau}}$. It is clear that as $\tau \rightarrow \infty$, i.e., using more local data to compute the demand estimate, the metric used in the case of LRFU approaches that of the optimal cache hit in \eqref{eq:the_problem}. The above result is independent of the demand process, as opposed to the existing work on LRFU, which typically assume i.i.d. demands. The following section presents simulation results to validate some of the insights provided by our theory to design online caching algorithm, and compare it with some of the well known algorithms. 

\section{Simulation Results}
\begin{figure}[h!]
    \centering
    \includegraphics[width=19em,height=18em]{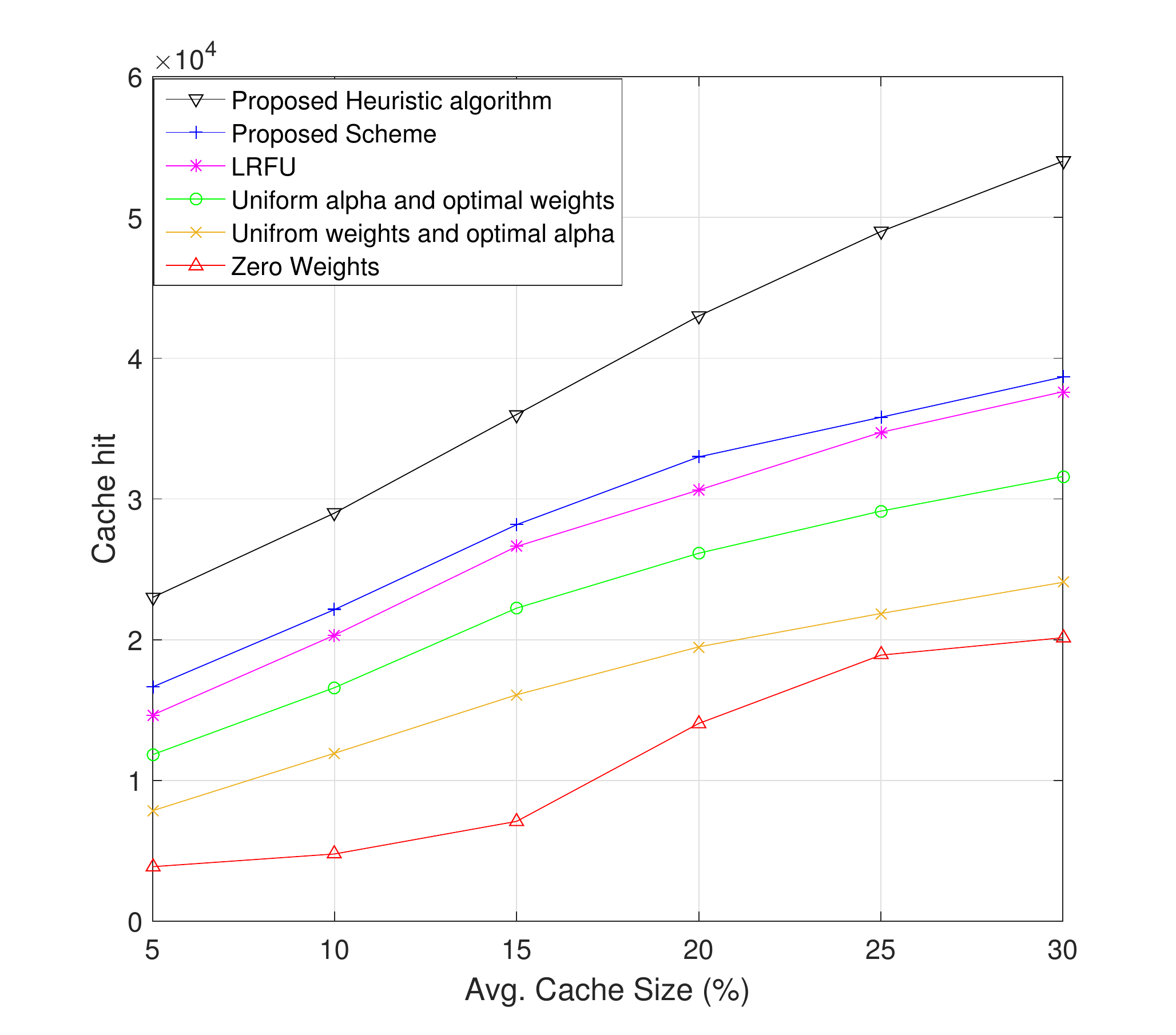}
    \caption{Average cache hit versus cache size for $1$st sBS.}
    \label{fig:cache_hit_sBS1}
\end{figure}

\begin{figure}[h!]
    \centering
    \includegraphics[width=19em,height=18em]{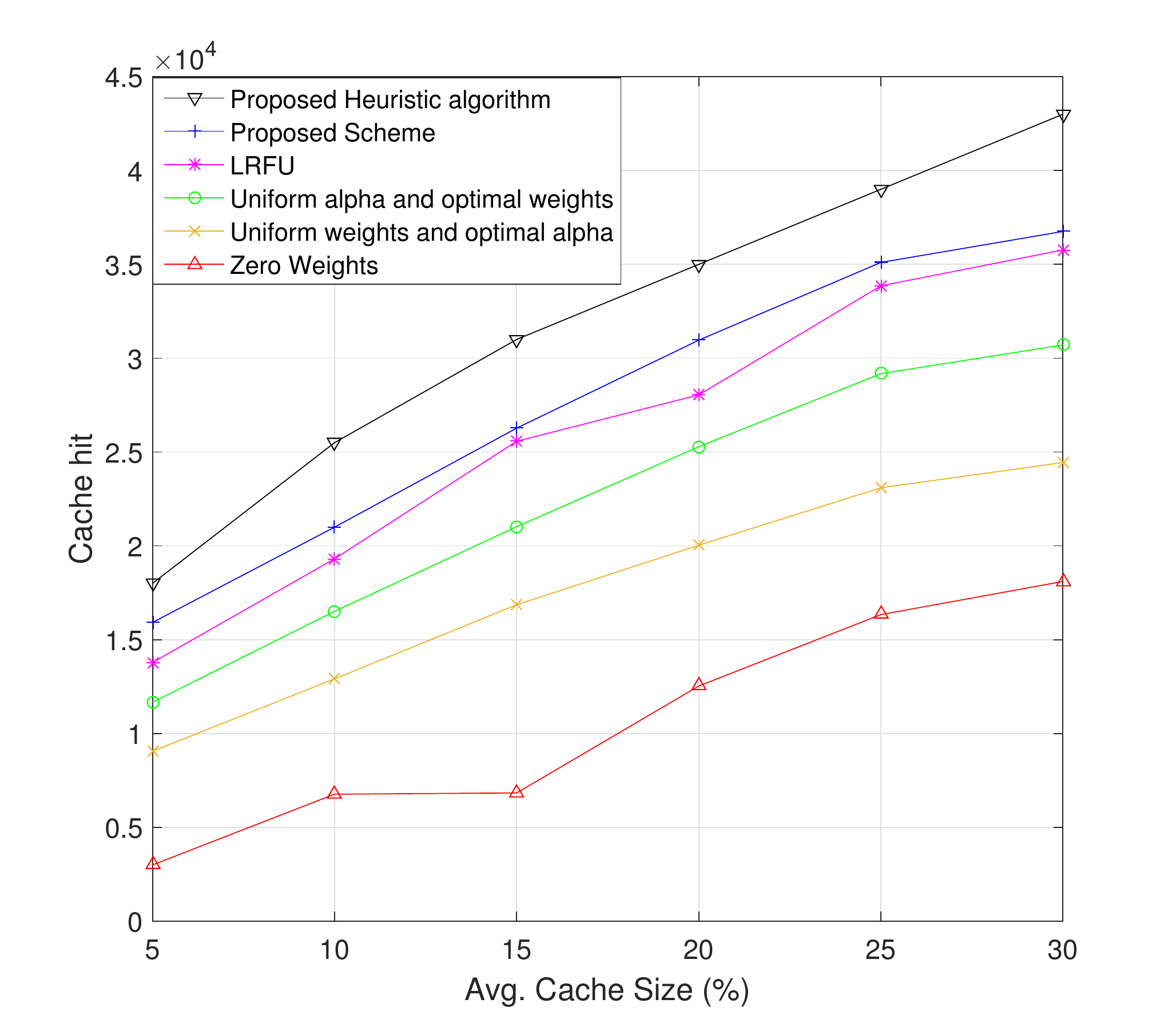}
    \caption{Average cache hit versus cache size for $5$th sBS.}
    \label{fig:cache_hit_sBS5}
\end{figure}

\begin{figure}[h!]
    \centering
    \includegraphics[width=19em,height=18em]{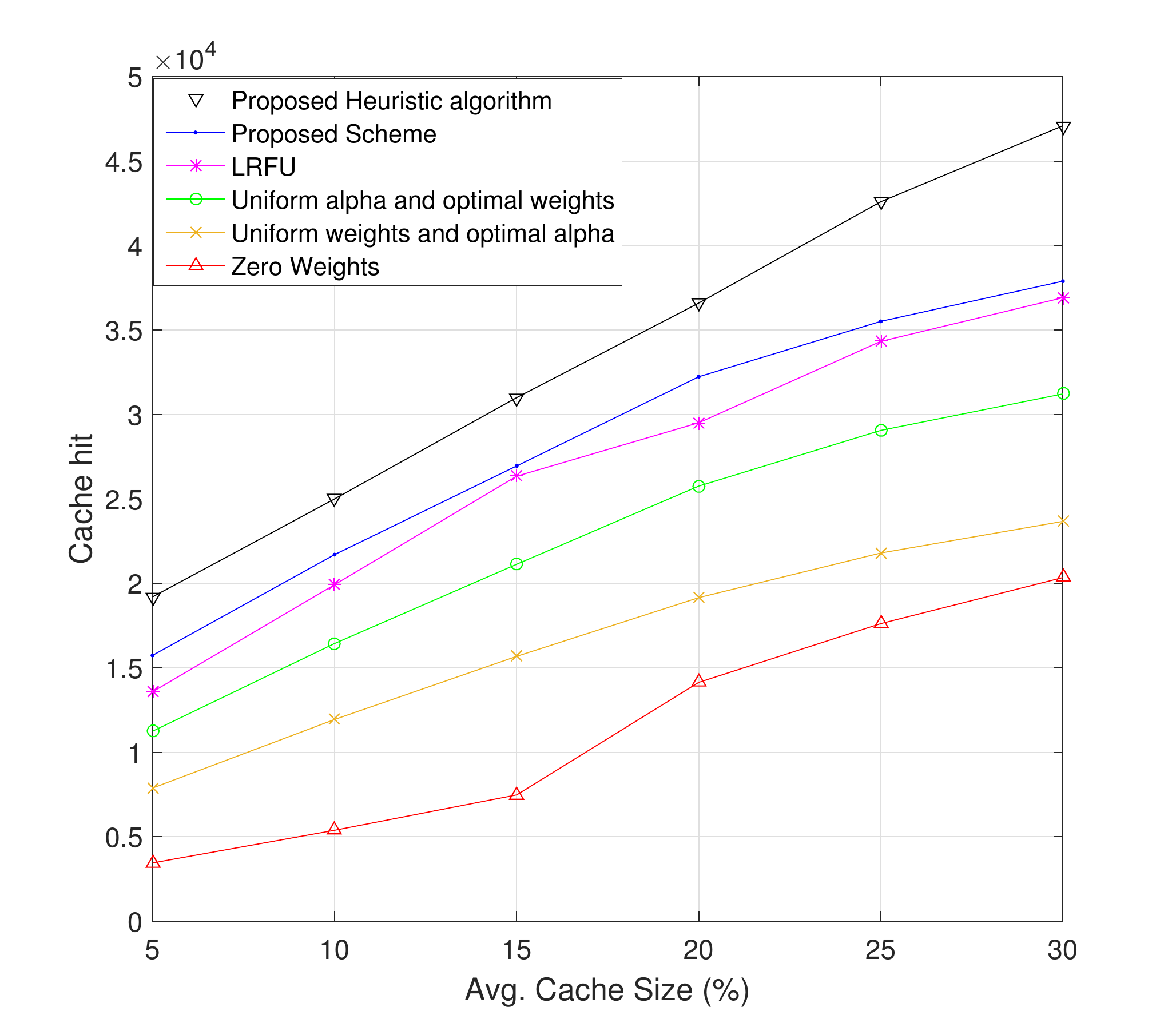}
    \caption{Average sum cache hit of all sBSs versus cache size.}
    \label{fig:sum_cache_hit}
\end{figure}

\begin{figure}[h!]
	\centering
	\includegraphics[width=19em,height=18em]{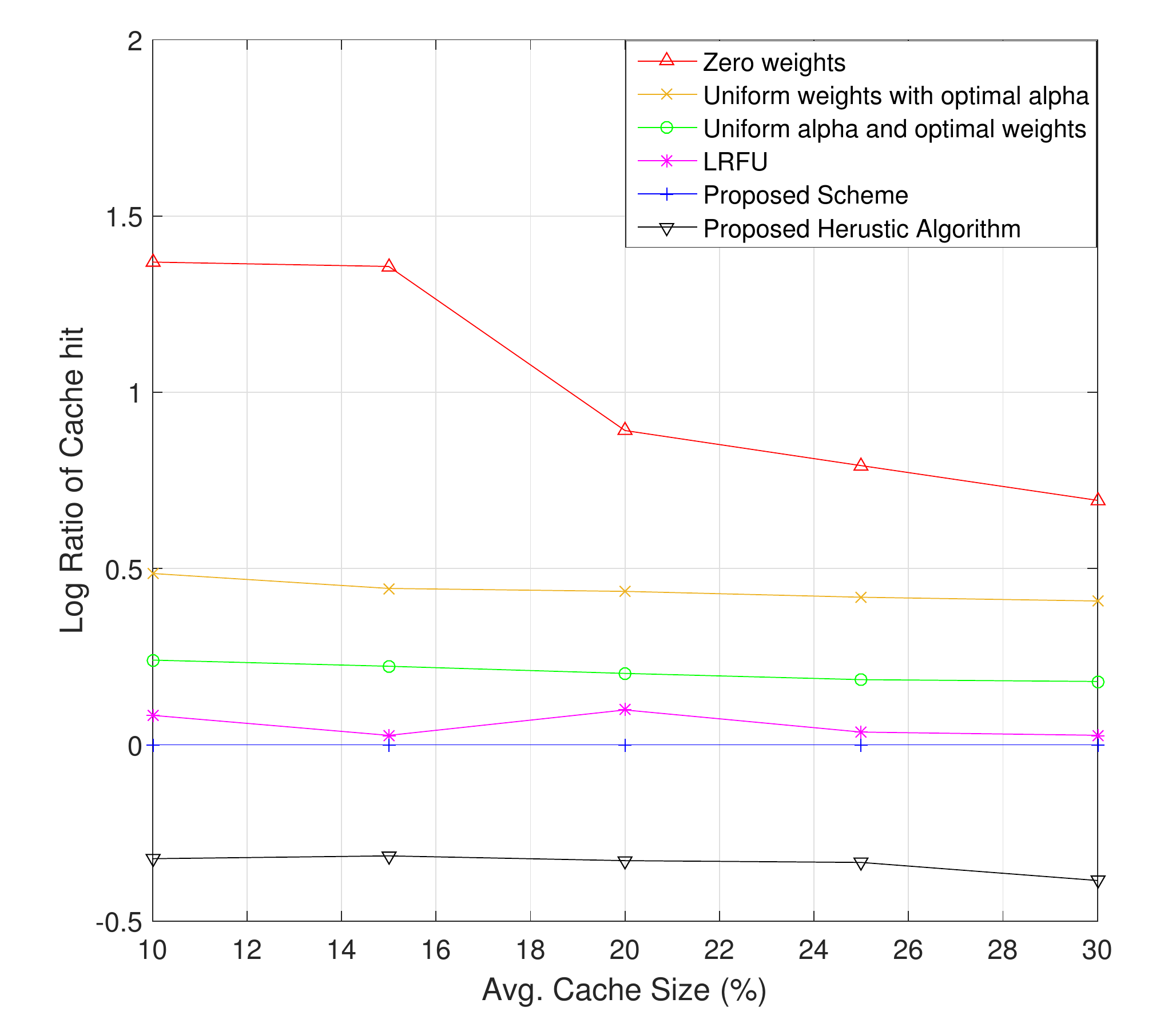}
	\caption{Log ratio of cache hit of average of all sBSs versus cache size.}
	\label{fig:log_diff_cache_hit}
\end{figure}

\begin{figure}[h!]
	\centering
	\includegraphics[width=19em,height=18em]{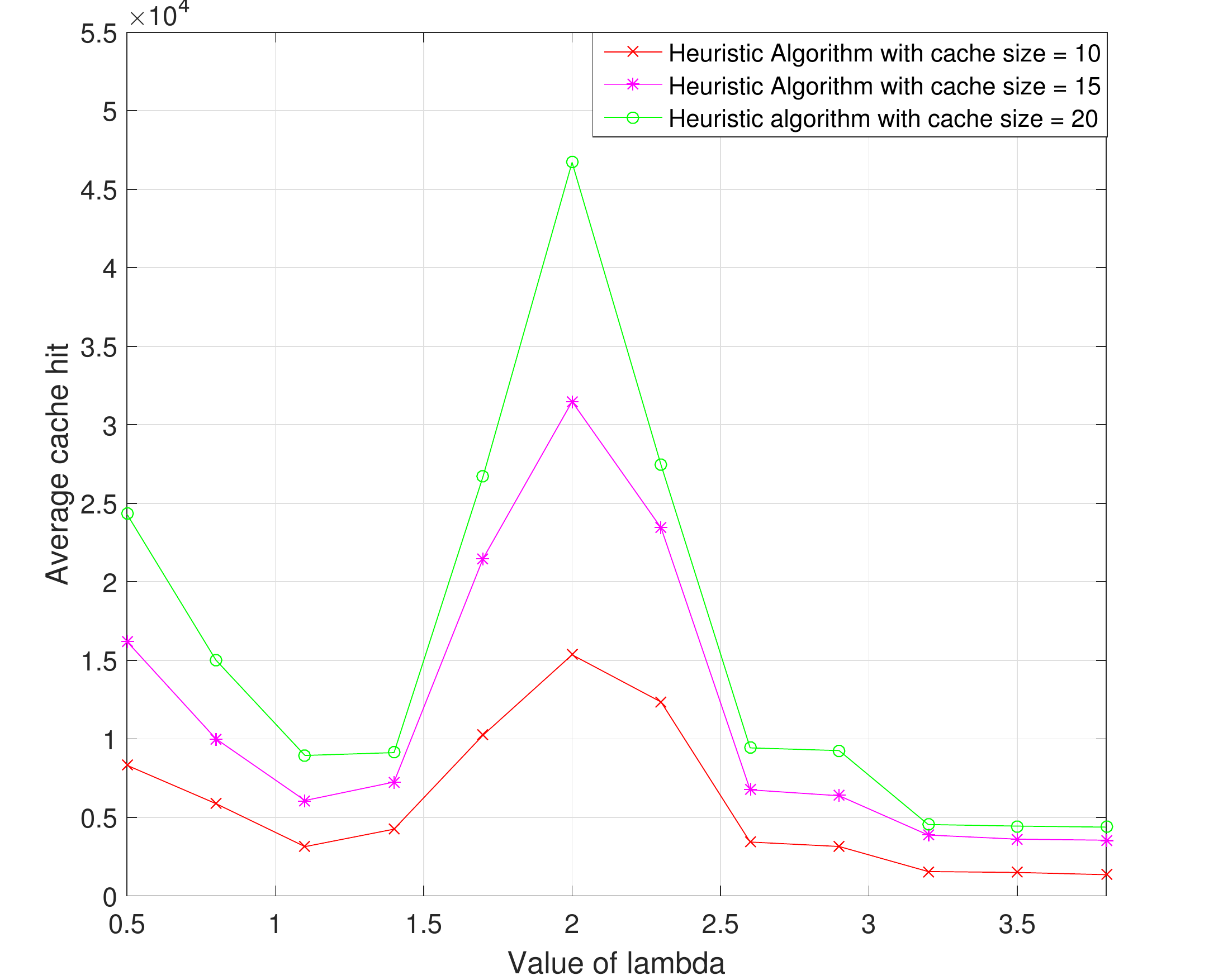}
	\caption{Average cache hit of Heuristic algorithm versus $\lambda$ for different cache sizes.}
	\label{fig:lambda_var}
\end{figure}

The simulation setup consists of five sBSs with several users connected to each of the sBS as shown in Fig.~\ref{fig:caching_EC}. The topology of the sBSs are described by $1 \leftrightarrow 2 \leftrightarrow 3$, $3 \leftrightarrow 4 \leftrightarrow 5$, and $5 \leftrightarrow 1$, where $a \leftrightarrow b$ indicates that sBSs $a$ and $b$ can communicate with each other. Without loss of generality, it is assumed that the users can move, and over time connect to different sBSs. The demands from the users is generated using the Movie Lens data set.\footnote{http://grouplens.org/datasets/movielens/} The total number of files is $800$, i.e., the users can possibly request from only these catalog of MovieLens data. The size of each file is assumed to be chosen uniformly random from $10$ to $100$ units. The demands at each sBS is obtained by randomly dividing Movie Lens data into $5$ disjoint chunks, which is spread across $200$ time slots. Further, the demands are normalized in each slot to get the popularity profile. This is used in place of demands while defining the (weighted cache hit and discrepancy) metric to compute the optimal weights in \textbf{Algorithm $\bm{1}$}. 
The average cache hit with un-normalized demands is used as a performance measure. The optimization is done with respect to the weights across time as well as sBSs. In this section, for simplicity, the weights across time will be referred to as $\alpha$, and the weights allocated across sBSs as $\bm{w}$. To understand the importance of past demands and the neighboring sBSs demand, it is important to compare the proposed scheme under various conditions. In particular, the proposed algorithm is compared with (i) the heuristic algorithm proposed in Sec.~\ref{sec:heuristic}, (ii) the algorithm that uses uniform $\bm{w}$ and optimal $\alpha$, (iii) LRFU, (iv) algorithm with uniform $\alpha$ and optimal $\bm{w}$, and (v) algorithm with optimal $\alpha$ and $\bm{w}=0$ resulting in zero weights (i.e the neighboring sBSs data is not used in the caching policy). The following parameters were used: $\tau = 10$, $\tau_1 = \tau_2 = 5$, $\eta_k = 1/\sqrt{k}$, $\beta_k = 0.01/\sqrt{k}$, and $\gamma_k = 0.4/\sqrt{k}$, where $k$ is the iteration index in the algorithm. Figs.~\ref{fig:cache_hit_sBS1} and \ref{fig:cache_hit_sBS5} show the average cache hit versus cache size as a fraction of the total size of the catalog for sBSs $1$ and $5$, respectively. It is clear from the figure that the proposed algorithm (both proposed weighted averaging caching algorithm and proposed heuristic caching algorithm) performs better than the LRFU, optimal $\alpha$, and $\bm{w}=0$, uniform $\alpha$ and optimal $\bm{w}$, as well as uniform $\bm{w}$ with optimal values of $\alpha$. The difference in the average cache hit is of the order of $10^4$ for both sBS $1$ and sBS $5$. Fig.~\ref{fig:sum_cache_hit} shows the sum cache hit rate of all the sBSs summarizing the trends in all the sBSs. The difference here is around $10^4$ demonstrating the benefit of using the proposed scheme(s). Fig.~\ref{fig:log_diff_cache_hit} shows the logarithmic ratio of the average cache hit of the proposed scheme with all other algorithms. This is done to show the gaphs more clearly. Since, only the heuristic algorithm performs better than the proposed scheme, we can see that the ratio of the proposed scheme with that of the proposed heuristic algorithm will lead to a negative value, and it is positive for the remaining algorithms. The ratio of the proposed scheme with itself is one and since logarithm of one is zero, hence the ratio of the proposed scheme with itself will be zero. Fig.~\ref{fig:lambda_var} shows the variations of heuristic algorithm with respect to $\lambda$. It is observed that for $\lambda = 2$, the heuristic algorithm performs the best and hence this value of $\lambda = 2$ is chosen for comparison with the other algorithms. 


\section{Remarks and Future Directions}\label{sec:conclusion}
The paper proposed an algorithm for caching in a distributed cellular network setting using theoretical guarantees provided in Theorem \ref{thm:mainresult_codedcaching}. It is shown that using optimal weights obtained in the algorithm outperforms both LRFU and the algorithm with equal weights. The proposed algorithm, uses a discrepancy measure with the regret minimization. The LRFU algorithm uses a windowed average of demands, and caches the files with the highest average demands. Despite the simplicity of the algorithm, there are no theoretical guarantees when the demands are non-stationary and hence, theoretical guarantees on the performance of the LRFU caching strategy is provided in this work. Further, federated learning based heuristic caching algorithm is also proposed and it is observed that it performs better than the weighted caching algorithm and hence motivating the future work on providing guarantees for the heuristic algorithm. Finally, it is interesting to explore average weighted demands in place of average demands in the LRFU performs better than the vanilla LRFU and the proposed algorithm. In this case, how should one choose the weights? Answers to these questions will be a part of our future work.

\section{Proof of Theorem \ref{thm:mainresult_codedcaching}} \label{sec:proof_main_res_codedcaching}
Assume that each SBS $b$ employs the caching strategy in \eqref{eq:pistart_coded_caching} based on the local data $Z_{b,1}^T$. Then, the corresponding conditional average of the hit rate is given by
\begin{eqnarray}
\mathbb{E}\left[\mathcal{R}_{b,T+1}(\bm{\pi_{b,T+1}}^{(av)}) \left | \right. Z_{b,1}^T\right] &\stackrel{(a)}{=}&   w_{b}^{T+1} \sum_{t=T-\tau}^T \alpha_{b,t} \mathbb{E}\left[\mathcal{R}_{b,T+1}(\bm{\pi}_{b,t}) \left | \right. Z_{b,1}^T\right] + \nonumber \\
&& \sum_{b^{'} \in \mathcal{N}_b} w_{b^{'}}^{T+1} \sum_{t=T-\tau}^T \alpha_{b^{'},t} \mathbb{E}\left[\mathcal{R}_{b,T+1}(\bm{\pi}_{b^{'},t}) \left | \right. Z_{b,1}^T\right] \nonumber\\
&\stackrel{(b)}{=}&  \sum_{t=T-\tau}^T \alpha_{b,t} \mathbb{E}\left[\mathcal{R}_{b,T+1}(\bm{\pi}_{b,t}) \left | \right. Z_{b,1}^T\right] - \texttt{M}_{b,T+1}(\bm{w}_{\neq b, T}), \label{eq:first_equality_coded_caching}
\end{eqnarray}
where $(a)$ follows simply by substituting for $\bm{\pi_{b,T+1}}^*$ from \eqref{eq:pistart_coded_caching}. The equality $(b)$ follows by (i) adding and subtracting the term $\sum_{b^{'} \in \mathcal{N}_b} w_{b^{'}}^{T+1} \sum_{t=T-\tau}^T \alpha_{b,t} \mathbb{E}\left[\mathcal{R}_{b,T+1}(\bm{\pi}_{b,t}) \left | \right. Z_{b,1}^T\right]$, and using the definition of $\texttt{M}_{b,T+1}(\bm{w}_{\neq b, T})$, and (ii) using the fact that $w_{b}^{T+1} + \sum_{b^{'} \in \mathcal{N}_b} w_{b^{'}}^{T+1} = 1$ $\forall$ $b \in \mathbb{B}$. Now, by adding and subtracting $\sum_{t=T-\tau}^T\alpha_{b,t} \mathbb{E}\left[\mathcal{R}_{b,T+1}(\bm{\pi}_{b,t}) \left | \right. Z_{b,1}^{t-1} \right] $, and using the definition of $\mathbb{D}_{b,T}(\bm{\alpha_{b,T}})$ in \eqref{eq:disclg}, the above equation can be lower bounded as 
\begin{equation} \label{eq:lowerbound_deterministic}
\mathbb{E}\left[\mathcal{R}_{b,T+1}(\bm{\pi_{b,T+1}}^{(av)}) \left | \right. Z_{b,1
}^T\right]  \geq  \sum_{t=T-\tau}^T \alpha_{b,t}\mathbb{E}\left[\mathcal{R}_{b,t}(\bm{\pi}_{b,t}) \left | \right. Z_{b,1}^{t-1}\right] 
- \texttt{M}_{b,T+1}(\bm{w}_{\neq b, T}) - \mathbb{D}_{b,T}(\bm{\alpha_{b,T}}). 
\end{equation} 
Similarly, an upper bound can also be obtained as follows
\begin{equation} \label{eq:lowerbound_deterministic2}
 \mathbb{E}\left[\mathcal{R}_{b,T+1}(\bm{\pi_{b,T+1}}^{(av)}) \left | \right. Z_{b,1}^T\right]  \leq  \sum_{t=T-\tau}^T \alpha_{b,t} \mathbb{E}\left[\mathcal{R}_{b,t}(\bm{\pi}_{b,t}) \left | \right. Z_{b,1}^{t-1}\right] + \texttt{M}_{b,T+1}(\bm{w}_{\neq b, T}) + \mathbb{D}_{b,T}(\bm{\alpha_{b,T}})
\end{equation} 
where the above upper bound follows by adding the discrepancies instead of subtraction. Note that the term $$A_t :=   \alpha_{b,t} \mathcal{R}_{b,t}(\bm{\pi}_{b,t})  - \alpha_{b,t} \mathbb{E}\left[\mathcal{R}_{b,t}(\bm{\pi}_{b,t}) \left | \right. Z_{b,1}^{t}\right]$$ is a Martingale difference, i.e., $\mathbb{E}\left\{A_t \left | \right. Z_{b,1}^{t}\right\} = 0$. Thus, the following event occurs with a probability of at least $1-\delta$, which follows from the Azuma's inequality 
\begin{equation} 
\sum_{t=T-\tau}^T A_t \leq H_{\texttt{max}} \norm{\bm{\alpha_{b,T}}}_2 \sqrt{\frac{2}{\tau} \log\frac{1}{\delta}} .
\end{equation}
The above implies that
\begin{equation} \label{eq:azuma_At_firstequation}
 \sum_{t=T-\tau}^T \alpha_{b,t} \mathbb{E}\left[\mathcal{R}_{b,t}(\bm{\pi}_{b,t}) \left | \right. Z_{b,1}^{t-1}\right] \geq \sum_{t=T-\tau}^T  \alpha_{b,t} \mathcal{R}_{b,t}(\bm{\pi}_{b,t}) - H_{\texttt{max}} \norm{\bm{\alpha_{b,T}}}_2 \sqrt{\frac{2}{\tau} \log\frac{1}{\delta}},
\end{equation}
where $H_{\texttt{max}}$ is the maximum possible hit rate. Since $-A_t$ is also a Martingale difference, using Azuma's inequality, the following holds good with a probability of at least $1-\delta$
\begin{equation} \label{eq:azuma_At_secondequation}
\sum_{t=T-\tau}^T \alpha_{b,t} \mathcal{R}_{b,t}(\bm{\pi}_{b,t}) \geq  \sum_{t=T-\tau}^T  \alpha_{b,t} \mathbb{E}\left[\mathcal{R}_{b,t}(\bm{\pi}_{b,t}) \left | \right. Z_{b,1}^{t-1}\right] - H_{\texttt{max}} \norm{\bm{\alpha_{b,T}}}_2 \sqrt{\frac{2}{\tau} \log\frac{1}{\delta}}
\end{equation}
Using \eqref{eq:azuma_At_firstequation} in \eqref{eq:lowerbound_deterministic}, the following holds good with a probability of at least $1-\delta$
\begin{eqnarray} 
 \mathbb{E}\left[\mathcal{R}_{b,T+1}(\bm{\pi_{b,T+1}}^{(av)}) \left | \right. Z_{b,1}^T\right]  &\geq& \sum_{t=T-\tau}^T\alpha_{b,t} \mathcal{R}_{b,t}(\bm{\pi}_{b,t}) - H_{\texttt{max}} \norm{\bm{\alpha_{b,T}}}_2 \sqrt{\frac{2}{\tau} \log\frac{1}{\delta}} \nonumber\\
&&- \texttt{M}_{b,T+1}(\bm{w}_{\neq b, T})  - \mathbb{D}_{b,T}(\bm{\alpha}_{b,T}). \label{eq:azuma_At_thirdequation}
\end{eqnarray}
This proves the first result in the theorem. Similar to the above equation, using \eqref{eq:azuma_At_secondequation} in \eqref{eq:lowerbound_deterministic2}, the following holds good with a probability of at least $1-\delta$
\begin{eqnarray}
\sum_{t=T-\tau}^T \alpha_{b,t} \mathcal{R}_{b,T+1}(\bm{\pi}_{b,t}) &\geq& \mathbb{E}\left[\mathcal{R}_{b,T+1}(\bm{\pi_{b,T+1}}^{(av)}) \left | \right. Z_{b,1}^T\right]  - H_{\texttt{max}} \norm{\bm{\alpha_{b,T}}}_2 \sqrt{\frac{2}{\tau} \log\frac{1}{\delta}} \nonumber\\
&&- \texttt{M}_{b,T+1}(\bm{w}_{\neq b, T})  - \mathbb{D}_{b,T}(\bm{\alpha}_{b,T}) \label{eq:azuma_At_fourthequation}
\end{eqnarray}
Let $\bm{C}_{b,t}^*$, $t=T-\tau,\ldots,T$, $b \in \mathbb{B}$ be some sequence of caching strategy. 
Now, consider the following term
\begin{eqnarray} \label{eq:azuma_At_fifthequation}
-\sum_{t=T-\tau}^T \alpha_{b,t} \mathcal{R}_{b,T+1}(\bm{\pi}_{b,t}) + \sum_{t=T-\tau}^T \alpha_{b,t} \mathcal{R}_{b,T+1}(\bm{C}^*_{b,t}) &\leq& \sum_{t=T-\tau}^T \left(\alpha_{b,t} - \frac{1}{\tau}\right)\left(\mathcal{R}_{b,T+1}(\bm{C}^*_{b,t}) - \mathcal{R}_{b,T+1}(\bm{\pi}_{b,t})\right)  \nonumber \\
&& + \frac{1}{\tau} \sum_{t=T-\tau}^T\left(\mathcal{R}_{b,T+1}(\bm{C}^*_{b,t})  - \mathcal{R}_{b,T+1}(\bm{\pi}_{b,t})\right)   \nonumber \\
&\leq& H_{\texttt{max}} \sum_{t=T-\tau}^T \abs{\alpha_{b,t} - \frac{1}{\tau}} + \frac{ \texttt{Reg}_{b,T,\tau}(\bm{\pi}_{b,t}) }{\tau},
\end{eqnarray}
where the regret is as defined in \eqref{eq:regret}. If the caching strategy used is $\bm{C}^*_{b,t}$, then, 
the above implies that 
\begin{equation}
   \sum_{t=T-\tau}^T \alpha_{b,t} \mathcal{R}_{b,T+1}(\bm{\pi}_{b,t}) \geq  \sum_{t=T-\tau}^T \alpha_{b,t} \mathcal{R}_{b,T+1}(C^{*}_{b,t}) - H_{max} \sum_{t=T-\tau}^T \abs{\alpha_{b,t}- \frac{1}{\tau}} - \frac{ \texttt{Reg}_{b,T,\tau}(\bm{\pi}_{b,t}) }{\tau}.
\end{equation}
From \eqref{eq:azuma_At_thirdequation}, we have 
\begin{eqnarray}
    \mathbb{E}[\mathcal{R}_{b,T+1}(\bm{\pi_{b,T+1}}^{(av)})|Z_{b,1}^T] &\geq& \sum_{t=T-\tau}^T \alpha_{b,t} \mathcal{R}_{b,T+1}(\bm{C}^*_{b,t}) - H_{max} \sum_{t=T-\tau}^T \abs{\alpha_{b,T}- \frac{1}{\tau}} - \frac{ \texttt{Reg}_{b,T,\tau}(\bm{\pi}_{b,t}) }{\tau} \nonumber \\ && - H_{\texttt{max}} \norm{\bm{\alpha_{b,T}}}_2 \sqrt{\frac{2}{\tau} \log\frac{1}{\delta}} -  \texttt{M}_{b,T+1}(\bm{w}_{\neq b, T})  - \mathbb{D}_{b,T}(\bm{\alpha}_{b,T,\tau}).
\end{eqnarray}
Now, using \eqref{eq:azuma_At_fourthequation} with $\bm{C}_{b,T+1}^{*}:= \sum_{t=1}^T \alpha_{b,t} \bm{C}^{(av)}_{b,t}$ in place of $\bm{\pi}_{b,T+1}^{(av)}$, we get 
\begin{eqnarray}
 \mathbb{E}[\mathcal{R}_{b,T+1}(\bm{\pi_{b,T+1}}^{(av)})|Z_{b,1}^T] &\geq& \mathbb{E}\left[\mathcal{R}_{b,T+1}(\bm{C}^*_{b,T+1})|Z_{b,1}^T\right] - H_{max} \sum_{t=T-\tau}^T \abs{\alpha_{b,t}- \frac{1}{\tau}} - \frac{ 2\texttt{Reg}_{b,T,\tau}(\bm{\pi}_{b,t}) }{\tau}  \nonumber \\ &&- 2 H_{\texttt{max}} \norm{\bm{\alpha_{b,T}}}_2 \sqrt{\frac{2}{\tau} \log\frac{1}{\delta}} -  \texttt{M}_{b,T+1}(\bm{w}_{\neq b, T})  - 2\mathbb{D}_{b,T}(\bm{\alpha}_{b,T,\tau}). 
\end{eqnarray}
It is possible to choose $\bm{C}_{b,t}^*$ in such as way that $$\mathbb{E}\left[\mathcal{R}_{b,T+1}(\bm{C}^*_{b,T+1}) \left | \right. Z_{b,1}^T\right]   \geq \sup_{\bm{h}_{b,t}} \sum_{t=T-\tau}^T \alpha_{b,t} \mathbb{E}\left[\mathcal{R}_{b,T+1}(\bm{h}_{b,t})\left | \right. Z_{b,1}^T\right]   - \gamma$$ for some $\gamma > 0.$ Using this in the above equation, and substituting the resulting equation in \eqref{eq:azuma_At_thirdequation} gives 
\begin{eqnarray}
 \mathbb{E}\left[\mathcal{R}_{b,T+1}(\bm{\pi_{b,T+1}}^{(av)}) \left | \right. Z_{b,1}^T\right]  &\geq& \sup_{\bm{\pi}_{b,t}} \sum_{t=T-\tau}^T \alpha_{b,t} \mathbb{E}\left[\mathcal{R}_{b,T+1}(\bm{\pi}_{b,t})\left | \right. Z_{b,1}^T\right]   - 2H_{\texttt{max}} \norm{\bm{\alpha_{b,T}}}_2 \sqrt{\frac{2}{\tau} \log\frac{1}{\delta}}  \nonumber\\
&&\hspace{-2.9cm} - \texttt{M}_{b,T+1}(\bm{w}_{\neq b, T}) -  \frac{2 \texttt{Reg}_{b,T,\tau}(\bm{\pi}_{b,t}) }{\tau}  - H_{\texttt{max}} \sum_{t=T-\tau}^T \abs{\alpha_{b,t}- \frac{1}{\tau}} -  2 \mathbb{D}_{b,T}(\bm{\alpha}_{b,T,\tau}) - \gamma.  \label{eq:azuma_At_sixthequation}
\end{eqnarray}

This completes the proof of the theorem. \qed

\section{Proof of Theorem \ref{thm:lrfu_guarantees}} \label{sec:proof_main_res_LRFU}
Note that the sequence $A_{b,s}:= \frac{1}{\tau} \left[\mathcal{R}_{b,s}(\bm{\pi}_{s}) -  \mathbb{E}\left\{\mathcal{R}_{b,s}(\bm{\pi}_{s}) | Z_{b,1}^{s-1} \right\}\right]$ for $t - \tau -1 \leq s \leq t-1$ is a Martingale difference. The sequence is also bounded, i.e., $\abs{A_{b,s}} \leq \frac{H_{\texttt{max}} \norm{\bm{\alpha_{b,T}}}_2}{\tau}$. Hence, by Azuma's inequality, it can be seen that with a probability of at least $1-\delta$, for any caching strategy $\bm{\pi}_b$, the following holds
\begin{equation}
\sum_{f} {\pi}_{b,t} \hat{d}_{b,f,t} \mathcal{L}_f \leq \frac{1}{\tau} \sum_{s = t - \tau -1}^{t-1}\mathbb{E}\left[\mathcal{R}_{b,s}(\bm{\pi_{b}}) \left | \right. Z_{b,1}^{s-1}\right] + H_{\texttt{max}} \norm{\bm{\alpha_{b,T}}}_2 \sqrt{\frac{2 \log{\frac{1}{\delta}}}{\tau}},
\end{equation}
where the estimate $\hat{d}_{b,f,t} := \frac{1}{\tau} \sum_{s=t-\tau - 1}^{t-1} d_{b,f,s}$ for all $f$. Now, the following bound can be obtained by adding and subtracting $\mathbb{E}\left[\mathcal{R}_{b,t}(\bm{\pi_{b}}) \left | \right. Z_{b,1}^{t-1}\right]$, and taking the supremum of the modulus over caching strategies to get the following bound in terms of discrepancy 
\begin{equation}
\sum_{f} \bm{\pi}_{b,t} \hat{d}_{b,f,t} \mathcal{L}_f \leq \mathbb{E}\left[\mathcal{R}_{b,t}(\bm{\pi_{b}}) \left | \right. Z_{b,1}^{t-1}\right] + \mathbb{D}_{b,t}(\bm{u}_{\tau}) + H_{\texttt{max}} \norm{\bm{\alpha_{b,T}}}_2 \sqrt{\frac{2 \log{\frac{1}{\delta}}}{\tau}}.
\end{equation}
Similarly, the following bound can be obtained by adding and subtracting $\mathbb{E}\left[\mathcal{R}_{b,t}(\bm{\pi_{b}}) \left | \right. Z_{G,1}^{t-1}\right] $, and taking supremum of the modulus over all caching strategies (as done previously) to get 
\begin{equation}
\sum_{f} \bm{\pi}_{b,t} \hat{d}_{b,f,t} \mathcal{L}_f \leq \mathbb{E}\left[\mathcal{R}_{b,t}(\bm{\pi_{b}}) \left | \right. Z_{G,1}^{t-1}\right]  + \mathbb{D}_{b,t}(\bm{u}_{\tau}) + \mathbb{D}_{GL,t}(\bm{\alpha}_b)  + H_{\texttt{max}} \norm{\bm{\alpha_{b,T}}}_2 \sqrt{\frac{2 \log{\frac{1}{\delta}}}{\tau}}.
\end{equation}
The desired result in the theorem can by obtained by taking supremum over all caching strategies $\bm{\pi}_b$, and identifying that the supremum in the right hand side results in the LRFU caching strategy. This completes the proof. \qed

\bibliographystyle{IEEEtran}
\bibliography{caching_bib}

\end{document}